\documentstyle[aps,preprint]{revtex}
\begin{document}
\preprint{submitted to Phys. Rev. B}
\draft
\title{Manifestation of Quantum Chaos in Electronic Band Structures}
\author{E. R. Mucciolo, R. B. Capaz, B. L. Altshuler,
and J. D. Joannopoulos}
\address{Department of Physics, Massachusetts Institute of Technology,
Cambridge, Massachusetts 02139, USA}
\date{April 5, 1994}
\maketitle

\begin{abstract}
We use semiconductors as an example to show that quantum chaos
manifests itself in the energy spectrum of crystals. We analyze the
{\it ab initio} band structure of silicon and the tight-binding
spectrum of the alloy $Al_xGa_{1-x}As$, and show that some of their
statistical properties obey the universal predictions of quantum chaos
derived from the theory of random matrices. Also, the Bloch momenta
are interpreted as external, tunable, parameters, acting on the
reduced (unit cell) Hamiltonian, in close analogy to Aharonov-Bohm
fluxes threading a torus. They are used in the investigation of the
parametric autocorrelator of crystal velocities. We find that our
results are in good agreement with the universal curves recently
proposed by Simons and coworkers.
\end{abstract}

\section{Introduction}

The great success of band structure theory in providing a very
accurate description of electronic properties of many materials is a
well-established fact. Using few ingredients (like crystal structure
and atomic number), one can obtain a great variety of
material-dependent results (optical and transport properties, phonon
spectrum, etc.) in very good agreement with the available experimental
data. However, the implementation of a realistic band structure
calculation is not a straightforward task and usually requires very
complicated numerical algorithms. One is then led to ask whether there
is some universal (material-independent) behavior which may lay hidden
in the apparent regularity of the bands. The universality should
express some common characteristic of the underlying physical systems,
in this case crystals. For instance, in the {\it muffin-tin
approximation}, the motion of a valence electron inside a crystal can
be pictured as that of a particle in a periodic billiard structure of
smooth walls, whose classical dynamics is very likely to be chaotic
\cite{laughlin87,gutzwiller90}. If this simple analogy is valid, then
one should be able to find in the electronic spectra of real crystals
some of the universal signatures of quantum chaos
\cite{haake91,bohigas91}.

Quantum chaos occurs when a system exhibits chaotic dynamics in the
classical limit. The best way to observe chaos is to break all
continuous symmetries, so that the only constant of motion left is the
total energy. In this case, we expect the energy spectrum to show some
chaotic behavior, as well as some regularities (the so-called clean
features). Chaos and discrete symmetries can coexist: one good example
is the Sinai billiard \cite{sinai70}. However, discrete symmetries
imply the partition of the eigenstates into classes and, consequently,
the existence of degeneracies in the spectrum. As we shall see below,
lifting these degeneracies usually means removing part of the clean
features, resulting in an enhancement of the universalities.

In a crystalline material the translation invariance allows us to
reduce the problem to the study of the electron motion in a single
unit cell with quasiperiodic boundary conditions. Denoting the
periodic part of the single-electron wave function as
$u_{\vec{k}}(\vec{r}\ )$, the Schr\"odinger equation becomes
\begin{equation}
H_{\vec{k}} \ u_{\vec{k}}(\vec{r}\ ) = E(\vec{k}) \
u_{\vec{k}}(\vec{r}\ ) \ ,
\label{eq:c1.1}
\end{equation}
where $\vec{k}$ is the Bloch momentum and the reduced Hamiltonian is
\begin{equation}
H_{\vec{k}} = -\frac{\hbar^2}{2m} ( \ \nabla + i \vec{k} \ )^2 +
V(\vec{r}\ ) \ .
\label{eq:c1.2}
\end{equation}
The effective potential $V(\vec{r}\ )$ has all the discrete symmetries
(rotation, inversion, reflection, etc.) of the unit cell. The
quasiperiodic boundary conditions provide a torus geometry to the unit
cell and the Bloch momenta act as Aharonov-Bohm fluxes \cite{berry86}.
Notice that for $\vec{k}=0$ (at the so-called $\Gamma$ point) the
Hamiltonian is real and time-reversal ({\it T }) symmetric. Any
$\vec{k}\ne 0$ internal to the Brillouin zone (i.e., which is not
equivalent to $-\vec{k}$ by umklapping) breaks {\it T}. The potential
removes the translational symmetry within the torus, making the
problem chaotic.

Normally, one solves Eq.~(\ref{eq:c1.1}) for $\vec{k}$ varying along
some special symmetry lines of the Brillouin zone and most bands
calculated in this way will be degenerate. When we vary $\vec{k}$ away
from the planes and lines of symmetry, all discrete symmetries of
$H_{\vec{k}}$ are broken and all degeneracies are lifted, resulting in
a ``spaghetti'' of bands (in this work we do not consider spin-orbit
coupling or strong magnetic fields and Kramers degeneracy is {\it
always} present). As an illustration, suppose we select a set of bands
which are degenerate at the $\Gamma$ point and then follow them as we
move in $\vec{k}$-space. We will notice that the bands split;
eventually, some of them may come close together again, but if we do
not cross any symmetry line, they will always ``repel'' (anticross)
each other. Therefore, the Bloch momentum can be regarded as an
external parameter which drives the system out of the partial
integrability of the symmetry points and into a region where the
spectrum is chaotic and bands are strongly correlated.

We expect the regions of the spectrum where chaos dominates to be
short-ranged, since chaos and the consequent universality appear at
very long time scales, translating into short energy scales. The clean
features, on the other hand, are connected to the classical periodic
orbits (short time scale) \cite{gutzwiller90}, which do not probe
large portions of phase space: They will prevail in regions of the
Brillouin zone close to the symmetry points, but can also be visible
elsewhere.

We should contrast this view of chaos in crystals whose unit cells are
fairly simple (few atoms per unit cell) with the more commonly
discussed case of chaos in small disordered metallic grains (quantum
dots). When there is no order or symmetry, but the system is small
enough for individual energy levels to be distinguishable, it is very
natural to study the level repulsion characteristic of quantum chaos
and consider its implications to the thermodynamical properties of the
system \cite{gorkov65,muehlschlegel91}. One can, for example, explore
how much disorder is necessary to switch the system from nonchaotic to
chaotic and how this crossover takes place \cite{bohigas86}. Another
idea which so far has only been developed theoretically is the study
of chaos in arrays of identical disordered unit cells
\cite{laughlin87,nobuhiko93}. The translational invariance in these
systems leads to the Hamiltonian of Eq.~(\ref{eq:c1.2}) and the same
considerations we have just made for an elementary crystal should
apply to the array of identical quantum dots. The only difference in
this case is that the unit cell has no discrete symmetry and the
spectrum is more likely to indicate strong chaotic dynamics. The clean
features which persist in the band structure of the supercrystal will
disappear after ensemble averaging.

The traditional diagnostic of quantum chaos derives from the original
works of Wigner, Dyson, and Mehta on the theory of random matrices
\cite{mehta91}, which was initially designed for the study of
statistical properties in Nuclear Physics. They have introduced most
of the necessary mathematical tools: level spacing distributions,
cluster functions, and several statistics, all focusing on level
correlations. When a quantum system is subjected to an external
perturbation, there are alternative ways to characterize chaoticity
\cite{haake91,wilkinson90,gaspard90,delande93,goldberg91,aaron93,ben93,beenakker93}.
The parametric correlation functions
\cite{goldberg91,aaron93,ben93,beenakker93} obtained by this
approach are directly related to the transport properties of the
system, hence bringing some important physical insight.

In this paper we show that an appropriate analysis of energy spectra
obtained by band structure calculations indicates the unambiguous
manifestation of quantum chaos in crystalline materials. Here we
consider $Si$ as an example of a crystal which can be viewed as a
quantum chaotic system with a particularly simple (diatomic) unit
cell. Despite the fact that {\it T} is broken for an internal
$\vec{k}$, the space-inversion symmetry of the $Si$ crystal yields a
{\it false} time-reversal violation \cite{robnik86} and the system is
described by the Gaussian orthogonal ensemble (GOE) in the entire
Brillouin zone. In contrast to $Si$, we also study the supercrystal
formed by complex $Al_xGa_{1-x}As$ cells. In this case chaos can be
enhanced by increasing the amount of disorder through changing the
concentration $x$. The lack of inversion symmetry of the unit cell
causes {\it T} to be quickly broken outside the $\Gamma$ point and
therefore the ensemble is unitary (GUE).

\section{The Silicon Band Structure and Quantum Chaos}

In order to look for quantum chaos, we have to avoid doing the
analysis of the spectrum at the $\Gamma$ point or at any symmetry
point of the Brillouin zone. The effect of setting an internal
$\vec{k}\ne 0$ is not just to lift the degeneracies caused by the
point-symmetries of the unit cell. The Bloch momentum also serves to
help increase the statistics by acting as a three-component external
parameter. Once we have reached a region in the Brillouin zone where
chaos is well developed and regularities are weak, we can tune
$\vec{k}$ to generate a large number of spectra and hence facilitate
the analysis.

As we have argued in the Introduction, even crystals with fairly
simple unit cells should show quantum chaos. In order to verify this
prediction, we have performed {\it ab initio} electronic structure
calculations of the $Si$ energy bands. Our calculations were based on
the local-density-functional and pseudopotential approximations.
Details of the method are presented elsewhere \cite{payne92}. A
plane-wave cutoff of 15 Rydbergs was used in order to ensure a
faithful description of the higher bands, which are most likely to
show chaotic behavior than the low-lying ones.

The typical band dispersion for $Si$ is shown in Figure 1. The
momentum varies from the $\Gamma$ point to the boundary of the
Brillouin zone, passing by the center of mass (CM) of the irreducible
part. Notice the band splitting for $\vec{k}\ne 0$. If the bands are
truly uncorrelated, the distribution of band spacings $\epsilon$ is
Poisson-like \cite{haake91,bohigas91}: $P(\epsilon)
\propto e^{-\epsilon}$ (hereafter we will always express energies in
units of the mean band spacing). If any correlation is present, the
bands will tend to repel each other and $P(\epsilon\rightarrow
0)\rightarrow 0$.  In Fig.~1 we see that several sets of bands which
are degenerate at the $\Gamma$ point remain quite close together and
do not seem to interact very much.  This is caused by the presence of
symmetry lines in the vicinity of the direction we have chosen to plot
the bands.  Indeed, this ``memory effect'' is very strong in the band
structure of $Si$ and led us to concentrate our analysis at a small
region around the CM, which is reasonably far from the $\Gamma$ point
and other symmetry points.

It is clear from Eq.~(\ref{eq:c1.2}) that $H_{\vec{k}}$ is neither
invariant under {\it T}, nor under space inversion ({\it P}), in spite
of the fact that the potential $V(\vec{r}\ )=V(-\vec{r}\ )$ for $Si$.
In this case, however, $H_{\vec{k}}$ is invariant under the
antiunitary combination {\it TP} and this is sufficient to lead to GOE
fluctuations, instead of GUE as one might naively expect
\cite{robnik86}.

The two most popular diagnostics of quantum chaos originally from
random matrix theory (RMT) are the nearest-neighbor level spacing
distribution, $P(\varepsilon)$, and the rigidity of the spectrum, the
so-called $\Delta_3$ statistics. There is no expression for
$P(\varepsilon)$ in closed form, but, as an excellent approximation,
Wigner has proposed the surmise $P(\varepsilon) \propto
\varepsilon^\beta e^{-c_\beta \varepsilon^2}$, where $\beta=1$ and
$c_1=\pi/4$ for GOE, and $\beta=2$ and $c_2=4/\pi$ for GUE.  The
$\Delta_3$ statistics measures the variance of the number of levels
found in an interval of length $L$ \cite{mehta91}:
\begin{equation}
\Delta_3(L) = \frac{1}{L} \left\langle \mbox{Min}_{a,b}
\int_{\bar{E}-L/2}^{\bar{E}+L/2} dE \ [ N(E) - aE + b]^2
\right\rangle \ ,
\label{eq:c2.2}
\end{equation}
where $N(E)$ is the number of energy levels below the energy $E$. The
average indicated in Eq.~(\ref{eq:c2.2}) is performed over $\bar{E}$
(i.e., over nonoverlaping intervals between $\bar{E}-L/2$ and
$\bar{E}+L/2$), but in our study it is also taken over points in
$\vec{k}$-space. When the levels are completely uncorrelated (Poisson
statistics), we have $\Delta_3(L)=\frac{L}{15}$. In the opposite limit
of equally spaced levels, $\Delta_3(L)=\frac{1}{12}$. Sitting in
between these two limits are the curves drawn from RMT, which have the
$L\gg 1$ asymptotics
\cite{mehta91}
\begin{equation}
\Delta_3(L) \approx \frac{1}{\pi^2} \ln (L) - 0.00696
\ \ \ \ \ \ \mbox{(GOE)} \ ,
\label{eq:c2.3}
\end{equation}
and
\begin{equation}
\Delta_3(L) \approx \frac{1}{2\pi^2} \ln (L) + 0.0590
\ \ \ \ \ \ \mbox{(GUE)} \ .
\label{eq:c2.4}
\end{equation}

In addition to these two quantities, RMT has also a prediction about
the density-density (two-point) correlation function \cite{mehta91},
here defined as
\begin{equation}
R(\omega) = \langle \rho(\Omega + \omega) \rho (\Omega) \rangle -
\langle \rho \rangle^2 \ ,
\label{eq:c2.5}
\end{equation}
where $\rho(\Omega)=\sum_n \delta(\Omega-\varepsilon_n)$, and
$\varepsilon_n=E_n$. The function $R(\omega)$ behaves differently
depending on the particular ensemble: For GOE, it is linear close to
$\omega=0$, and then monotonically saturates to 1 at around
$\omega\approx$ 1, whereas for GUE it starts as quadratic and then
oscillates until it reaches saturation around the same values.

In Fig.~2a, 2b, and 2c we compare the statistical properties of the
$Si$ spectrum with the RMT predictions. The data was extracted from a
set of 80 high-energy eigenvalues corresponding to 343
$\vec{k}$-points (a $7\times7\times 7$ cube) around the CM. Notice the
good agreement with the GOE result, in contrast to GUE. The deviation
between the data points and the GOE curve for the $\Delta_3$
statistics at large $L$ (Fig.~2c) is expected because of the presence
of clean features when we consider large portions of the spectrum.

The Bloch momentum can be used as an external, continuous, parameter,
allowing us to evaluate ``dynamical'' universal correlation functions
of the spectrum. For this purpose, one needs a scaling parameter,
$\sqrt{C_{\mu\nu}(0)}$, which is related to the spectrum response to
$\vec{k}$; namely,
\begin{equation}
C_{\mu\nu}(0) = \left\langle \frac{\partial E_n(\vec{k})}{\partial
k_{\mu}} \frac{\partial E_n(\vec{k})}{\partial k_{\nu}} \right\rangle
\ ,
\label{eq:c2.6}
\end{equation}
whre the average is performed over many energy bands (the index $n$),
as well as over $\vec{k}$-points. Because in our study we dealt with
correlation over small regions in $\vec{k}$-space, it was a good
approximation to assume isotropy, i.e. $C_{\mu\nu}(0)=C(0)$ for all
$\mu,\nu$. After performing the rescaling $x=\sqrt{C(0)}\ k_\mu$,
where $\mu$ denotes some direction in $\vec{k}$-space, we evaluated
one of the simplest correlation function one can study
\cite{aaron93,ben93}, which is the autocorrelator of crystal
velocities,
\begin{equation}
c(x) = \left\langle \frac{\partial
\epsilon_n(\bar{x}+x)}{\partial{\bar{x}}}
\frac{\partial\epsilon_n(\bar{x})}{\partial{\bar{x}}}
\right\rangle \ ,
\label{eq:c2.7}
\end{equation}
It is important to notice that in both Eq.~(\ref{eq:c2.6}) and
(\ref{eq:c2.7}) the bands $E_n(\vec{k})$ have to be corrected for any
possible drift. The way we have proceeded was to estimate the local
average drift $\bar{v}_{n}=\langle \partial E_n(\vec{k})/\partial
\vec{k} \rangle$ and then subtract it from the crystal velocities.
We have found that $c(x)$ agrees reasonably with the universal form
introduced in Ref. \cite{ben93} for the pure GOE case (see Fig.~2d).
The small, but visible, discrepancy is understandable: We have used
only 7 points along each $\vec{k}$ direction and it is difficult to
perform a good estimate of $C(0)$ and $\bar{v}_n$ for such a short
interval.

\section{Quantum Chaos in the $A\lowercase{l}_{\lowercase{x}}
G\lowercase{a}_{1-\lowercase{x}} A\lowercase{s}$ alloy}

In contrast to crystalline materials with simple unit cells, where the
existence of symmetries causes partial integrability and the
regularities in the spectrum tend to hide the underlying chaotic
dynamics, disordered system are the ideal case to study. Classically,
the motion of an electrons inside a disordered grain is that of a
particle being repeatedly scattered by an irregular potential: The
complete lack of symmetries will give rise to a strong chaotic motion.
As a result, electronic disordered systems show very clear signatures
of quantum chaos and the concept of universality is generally valid.
The universal conductance fluctuations in mesoscopics systems
\cite{boris91} are a good example of a phenomenon related to the
chaotic dynamics of the electron in the sample.

If there is one disadvantage of disordered systems over pure
crystalline materials, it is the lack of Bloch momenta. From the
theoretical view point, the natural way to solve this problem is to
impose quasiperiodic boundary conditions to the disordered grain
\cite{nobuhiko93} and thus form a superlattice of identical complex
unit cells. The band structure of this supercrystal can then be
explored much in the same way as we did for $Si$.

In the analysis that follows we have chosen the widely studied
$Al_xGa_{1-x}As$ to demonstrate that it is indeed a good example of a
quantum chaotic system and to illustrate the applicability of the
parametric correlation functions to characterize quantum chaos. Alloys
are a good example of weakly disordered systems when their components
do not differ remarkably. If the sample is small enough (mesoscopic),
the average level spacing can be resolved experimentally
\cite{kastner92} and it makes sense to address the statistical
properties of the spectrum. The absence of discrete symmetries in the
unit cell guarantees that there are no degeneracies in the spectrum,
although some regularities may occurs and they are usually connected
to the nonuniversal features carried by the isolated components of the
alloy.

Our study of the $Al_xGa_{1-x}As$ supercrystals is based on a
semiempirical tight-binding method, with matrix elements taken from
the $sp^3s^{\ast}$ parametrization suggested by Vogl \cite{vogl83}.
Several ensemble realizations of a 216-atom basic cluster were
independently generated and solved (the electron wave function was
subjected to quasiperiodic boundary conditions). For each realization,
$Al$ and $Ga$ atoms were randomly distributed in the group-III
sublattice according to the aimed alloy composition. Alloy properties
are calculated as ensemble averages for each composition. This method
has been successfully used in the study of gap properties of the
random $Al_xGa_{1-x}As$ alloy \cite{rodrigo92}.

We begin by calculating the level spacing distribution of
$Al_xGa_{1-x}As$ at the $\Gamma$ point (hence the boundary conditions
are periodic) for three different compositions, namely $x\approx$ 0.1,
0.3, and 0.5 (note that for $0<x<1$ the spectrum is completely
nondegenerate). The results for the level spacing distribution are
shown in Fig.~3. For a given composition $x$, the averaging is done
over high-energy levels and over 20 realizations.  Notice that as the
disorder increases, we move from a Poisson-like law to a GOE-like,
Wigner-Dyson, distribution (remember that at the $\Gamma$ point there
is no {\it T} breaking). Another indication of the crossover between
weakly to strongly correlated energy levels can be seen in Fig.~4,
where we have plotted the $\Delta_3$ statistics for the three alloy
compositions.

Next we introduce phases to the boundary conditions to obtain the
dispersion of the bands with the Bloch momenta. In Fig.~5 we show a
typical set of bands at the high-energy part of the spectrum for a
composition $x\approx 0.5$. As for the case of $Si$, we restrict our
analysis to the region surrounding the CM. The clean features now are
fewer and usually related to regularities found in the band structure
of $AlAs$ and $GaAs$ \cite{folding}; one can be seen in the upper part
of Fig.~5.

Focusing on the high-energy bands and averaging over 9 realizations,
we obtain the results shown in Fig.~6. Notice the excellent agreement
with the GUE predictions. For any Bloch momentum outside the $\Gamma$
point, {\it T} is fully broken and since the unit cell is not
invariant under space inversion, we naively expect the statistics to
be GUE. In fact, this issue is more subtle. There is a continuous
change from GOE to GUE as we move away from the $\Gamma$ point
\cite{bohigas91} and the typical range of this crossover will depend on
the specific system under study. After the proper rescaling, the
crossover should be universal and at least two theoretical
investigations based on the supersymmetric technique have demonstrated
this point
\cite{altland93,nobuhiko94}. We have, as yet, not performed an
analysis of this crossover for the band structure of the
$Al_xGa_{1-x}As$ supercrystal. Nevertheless, we stress that our
results show that at the CM the system has already attained the GUE
limit.

It is worth to remark the good agreement between the data for $c(x)$
and the curve obtained in Ref. \cite{ben93} through numerical
simulations (see Fig.~6d). The deviation at $x\sim 1$ happens because
nonuniversal features dominate the correlation at large distances. We
mention that, in principle, we should be able to see the universal
behavior of the response function predicted in Ref. \cite{nobuhiko93}
by calculating suitable matrix elements of chaotic eigenstates. We
leave this subject for future investigation.

\section{Conclusions}

In summary, we have demonstrated that quantum chaos is present in the
band structures of $Si$ and of the $Al_xGa_{1-x}As$ supercrystal. We
have argued that this should hold true for any crystal because valence
electrons exhibit classical chaotic dynamics at the level of the unit
cell. The Bloch momentum can be viewed as an external parameter which
can be tuned to break the discrete symmetries of the unit cell and
revel the quantum chaos hidden in the regularity of the band
structure. Any crystal whose unit cell is invariant under space
inversion should be described by the GOE statistics. Violation of the
inversion symmetry combined with a deviation from the $\Gamma$ point
drives the system to GUE statistics. The parametric correlation
functions were shown to be a good diagnostic of quantum chaos, not
only for system with complex unit cells, like the supercrystal of
$Al_xGa_{1-x}As$, but also for the diatomic unit cell case of $Si$.
The important result of our work is that there exists universality in
band structures. The implications of this property to experiments may,
however, be limited. For instance, the optical properties of crystals
are usually defined by few, low-lying, bands close to the Fermi level,
which do not show very strong quantum chaos. One way to increase the
complexity of low energy levels is to consider real crystals with
polyatomic unit cells.

\begin{center}
{\large Acknowledgments}
\end{center}

The authors are greatful to Ben Simons for many illuminating
discussions throughout the evolution of this work. We also thank
Nobuhiko Taniguchi, Anton Andreev, and Tom\'as Arias for useful
conversations, and Belita Koiller for providing the tight-binding
code. E.R.M. and R.B.C. would like to acknowledge the financial
support of Conselho Nacional de Desenvolvimento Cient\a'{\i}fico
e Tecnol\'ogico (CNPq, Brazil). This work was supported by the Joint
Services Electronic Program no. DAAL 03-89-0001.

\figure{1. The band structure for $Si$. The Bloch momentum runs
from the $\Gamma$ point ($\vec{k}=0$) to the boundary of the Brillouin
zone, passing through the center of mass of the irreducible part
(CM). The scale is such that $k=1$ at the CM.
\label{fig1}}

\figure{2. Statistical properties of 90 high-energy bands of $Si$
around the CM point. The solid and dashed lines are the GOE and GUE
predictions, respectively. Figure (a) is the band spacing
distribution; (b) is the autocorrelator of density of states (rescaled
to $\langle \rho \rangle = 1$); (c) is the $\Delta_3$ statistics; (d)
is the crystal velocity correlation function $c(x)$. Error bars are of
the order of the data point symbol size or smaller, whenever not
indicated. For all curves the average was taken over the bands and
$\vec{k}$-points. The sets of points in (d) differ by the number of
bands used in the estimate of the local average drift,
$\langle\partial E_n(\vec{k}\ ) / \partial \vec{k} \rangle$. They
correspond to: 7 (circles), 9 (triangles), and 11 bands (diamonds).
\label{fig2}}

\figure{3. The level spacing distribution involving about 150
high-energy eigenstates at $\Gamma$ in the $Al_xGa_{1-x}As$
supercrystal for concentrations: (a) $x\approx 0.1$; (b) $x\approx
0.3$; (c) $x\approx 0.5$. The average was taken over 20 realizations
of a given concentration. The circles indicate the data and the solid
and dashed lines the GOE and Poisson statistics, respectively.
\label{fig3}}

\figure{4. The $\Delta_3$ statistics for the same set of levels and
realizations of Fig.~3, following the same conventions.
\label{fig4}}

\figure{5. Typical band dispersion around the CM point for the
supercrystal of $Al_xGa_{1-x}As$ cells ($x=0.5$). The region presented
here is particularly free of strong clean features, although a weak
one can be seen at the top of the spectrum. The scale of the momentum
axis has arbitrary units and is centered at the CM.
\label{fig5}}

\figure{6. Statistical properties for the high-energy bands of the
$Al_xGa_{1-x}As$ supercrystal around the CM point at $x=0.5$. The
circles indicate the data and the solid and dashed lines the GUE and
GOE predictions, respectively (error bars are typically smaller than
the circles). Figure (a) is the band spacing distribution; (b) is the
autocorrelator of density of states; (c) is the $\Delta_3$ statistics;
(d) is the crystal velocity correlation function $c(x)$. For (a), (b)
and (c) the average was taken over bands and ensemble (9
realizations); for (d) the average was only over ensembles. The data
for $c(x)$ was found to be insensitive to the way the average drift
was estimated.
\label{fig6}}

\begin{thebibliography}{10}
\bibitem{folding}
The band folding caused by the contraction of the Brillouin zone makes
it hard to associate a specific clean feature found in the band
structure of the $Al_xGa_{1-x}As$ supercrystal to its counterpart in
the band structures of $AlAs$ or $GaAs$.
\end{thebibliography}
\end{document}